\documentclass[sn-mathphys-num]{sn-jnl}
\usepackage{graphicx,times}%
\usepackage{multirow}%
\usepackage{amsmath,amssymb,amsfonts}%
\usepackage{amsthm}%
\usepackage{mathrsfs}%
\usepackage[title]{appendix}%
\usepackage{xcolor}%
\usepackage{textcomp}%
\usepackage{manyfoot}%
\usepackage{booktabs}%
\usepackage{algorithm}%
\usepackage{algorithmicx}%
\usepackage{algpseudocode}%
\usepackage{listings}%
\usepackage{color}
\newcommand{\MB}{}
\newcommand{\mga}{}
\newcommand{\VDR}{}
\begin{document}

\title[Article Title]{From Staging to Insight: An Educational Path to Understanding Bell’s Inequalities}

\author*[1,2]{\fnm{Valentina} \sur{De Renzi}}\email{vderenzi@unimore.it}

\author[3]{\fnm{Matteo G. A.} \sur{Paris}}\email{matteo.paris@fisica.unimi.it}

\author*[4]{\fnm{Maria} \sur{Bondani}}\email{maria.bondani@cnr.it}

\affil[1]{
\orgdiv{Dipartimento di Scienze Fisiche, Informatiche e Matematiche}, 
\orgname{University of Modena and Reggio Emilia},
 \postcode{41125}, 
 \country{Italy}}
\affil[2]{\orgdiv{Istituto Nanoscienze}, \orgname{National Research Council - CNR-NANO},   \postcode{I-41125}, \city{Modena}, \country{Italy}}
\affil[3]{
\orgdiv{Dipartimento di Fisica {\em Aldo Pontremoli}}, 
\orgname{Universit\'a degli Studi di Milano}, 
\postcode{I-20133}, 
\city{Milano},  
\country{Italy}}

\affil[4]{
\orgdiv{Institute for Photonics and Nanotechnologies}, 
\orgname{National Research Council - CNR-IFN}, 
 \postcode{I-22100}, \city{Como},
\country{Italy}}
\abstract{Quantum Physics is a cornerstone of modern science and technology, yet a comprehensive approach to integrating it into school curricula and communicating its foundations to policymakers, industrial stakeholders, and the general public has yet to be established. In this paper, 
 we discuss the rationale for introducing entanglement and Bell's inequalities to a non-expert audience, and how these topics have been presented in the exhibit  
“Dire l’indicibile” ("Speaking the Unspeakable"),  as a part of the Italian Quantum Weeks project. 
%This initiative aims to make quantum mechanics accessible to all, bridging the gap between complex scientific principles and public understanding. 
Our approach meets the challenge of simplifying quantum concepts without sacrificing their core meaning, specifically avoiding the risks of oversimplification and inaccuracy. Through interactive activities, including a card game 
demonstration and the VDR staging of CHSH experiments, participants explore the 
fundamental differences between classical and quantum probabilistic predictions. They gain 
insights into the significance of Bell's inequality verification experiments and the implications 
of the 2022 Nobel Prize in Physics. Preliminary results from both informal and formal 
assessment sessions are encouraging, suggesting the effectiveness of this approach.}
\maketitle
\section{Introduction}\label{sec1}
%%%
The story of the Einstein-Podolsky-Rosen (EPR) paradox \cite{EPR} and Bell's inequalities \cite{Bell,speakable} stands as a paradigmatic example in the history of 20th-century physics, 
one that  deserves communication to both students and the general public. This narrative begins 
with the achievements of quantum mechanics (QM), the most successful physical theory ever developed 
to describe the natural world, and the epistemological challenges it introduces. A central 
aspect of these challenges is the existence of entangled states, as formalized by the 
axiomatic framework of quantum mechanics completed in the 1930s.  

 In their seminal 1935 paper, Einstein, Podolsky, and Rosen \cite{EPR} describe the alleged paradox which derives from combining the `natural' assumption of reality and locality with the implications of QM rules, as applied  to highly non classical states - specifically, entangled states. The consequences of their {\it gedankenexperiment} suggest therefore that at least one of these assumptions must be false.    
%The {\it gedankenexperiment} proposed by Einstein, Podolsky, and Rosen in their seminal 1935 paper \cite{EPR} exemplifies the implications of applying quantum mechanics to highly nonclassical states—specifically, entangled states, alongside certain assumptions about the nature of reality and locality. By combining these assumptions, EPR constructed a paradox suggesting that at least one of them must be false. 

Their conclusion was that, as realism and locality should be considered as indispensable,  quantum mechanics must 
be incomplete and necessitating a broader theory to resolve its apparent paradoxes. The EPR 
problem, formulated in 1935, remained a theoretical exercise until the mid-1960s. During this 
period, intense debate arose over which assumption—realism, locality, or the completeness of 
quantum mechanics—should be abandoned. The de Broglie-Bohm hidden-variables theory emerged 
as a prominent attempt to provide a causal completion of quantum mechanics \cite{hidden1,hidden2}, while adherents of the Copenhagen interpretation chose to accept quantum theory as it stood,
effectively dismissing realism and/or locality as essential components of a factual description 
of nature. 

In 1964, John Bell reformulated the EPR problem by assuming the existence of hidden variables and derived constraints on the correlations between measurement outcomes performed on the two parts of a bipartite quantum state. These constraints, now known as Bell’s inequalities (BI), are of profound generality, relying solely on the assumptions of realism and locality. Crucially, they are violated when the bipartite state under consideration is entangled. This violation became a decisive experimental criterion for determining whether quantum physics could be considered a complete theory. The significance of Bell’s inequalities lies in their transformation of the metaphysical problem posed by EPR into a falsifiable mathematical statement, open to experimental verification. In the decades following Bell’s work, the central challenge shifted to designing reliable experiments capable of testing the violation of these inequalities. It was not until 2016 \cite{BBT} that experiments were conducted which closed all major experimental and conceptual loopholes, achieving universal acceptance. As a result, the violation of Bell’s inequalities has become one of the most rigorously tested and scrutinized classes of experiments in the history of physics.

The story of the alleged EPR paradox and Bell’s inequalities (BI) holds even greater significance today, as quantum physics, once a field confined to the foundations of physics, is now revolutionizing both technology and culture. The Second Quantum Revolution, fueled by profound advancements in our understanding and control of quantum systems, promises to transform industrial and societal landscapes. At the heart of this transformation lies the concept of entanglement, a cornerstone of quantum theory. The groundbreaking contributions to this field were recognized by the 2022 Nobel Prize in Physics, awarded to Alain Aspect, John F. Clauser, and Anton Zeilinger \cite{nobel} \emph{for their pioneering experiments with entangled photons, which confirmed the violation of Bell’s inequalities and laid the foundation for quantum information science}. Their work not only validated the fundamental principles of quantum physics but also provided a critical impetus for the development of quantum technologies, such as quantum computing and quantum cryptography, which rely on the unique properties of entangled particles.

We decided to make this story the core of our effort to communicate the concepts of Quantum Physics (QP)  towards high-school students and general public.  Though being aware of the inherent difficulty to grasp quantum concepts, the interactive exhibition,  ``Dire l'indicibile'' (``Speaking the unspeakable''), designed within the Italian Quantum Weeks project \cite{websiteIQWs}, strives to make QP comprehensible to everybody, using the natural language and examples from everyday experience.
This initiative  aims to bridge the gap between complex scientific truths and public understanding, accepting the challenge of simplifying quantum concepts without diluting their essence, avoiding the pitfalls of oversimplification and inaccuracy. The visitors of the exhibition are introduced to the basic concepts of QP, {\it i.e.}, definition of quantum states, superposition, quantum measurements, non commutativity of observables, entanglement,  through visual analogies and demonstrators.  Explicit reference to key results, such as Feynman's double slit and Stern-Gerlach experiments, are used to connect abstract concepts to physical evidence. Details of this path can be found in \cite{websiteIQWs} and will be published elsewhere \cite{manu}.

In the context of education and outreach activities triggered by the ongoing Second Quantum Revolution, several efforts have been made to update the way students and general public are introduced  to quantum concepts \cite{QTEdu,migdal2022visualizing,MDPI,10.1119/5.0211535}. Nevertheless, entanglement and its foundational relevance remain the most challenging concepts to communicate \cite{bitz,gili23}. Attempts have been made exploiting simulations and games \cite{10.1119/1.5086275,Lopez-Incera_2020,Foti,qplaylearn,QTris}, often avoiding any explicit reference to the mathematical formalism with the result of just giving 
suggestions and allusive explanations. In fact, aiming to communicate the concept of entanglement, one major challenge is the general widespread miscommunication about it, together with many other quantum concepts. For instance,  a critical point 
%in discussing entanglement 
is to clarify the differences between classical and quantum correlations, a distinction that is frequently overlooked or misrepresented.

In this paper, we focus specifically on  entanglement and Bell's inequalities, drawing from the recent Nobel Prize-winning research.  The paper is organized as follows: in Section~\ref{ent} we introduce the notion of entanglement, while Section~\ref{clas_corr_Bell} %provides a straightforward derivation 
presents  Bell’s inequalities in the CHSH form. Section~\ref{story} outlines our narrative approach to explaining entanglement, the EPR paradox, and Bell’s inequalities, while Sections~\ref{cards} and~\ref{stage} describe interactive activities demonstrating the CHSH inequality and recreating Nobel Prize-winning experiments. Section~\ref{validation} presents the educational implementation of this framework and its validation. Section~\ref{conc} 
closes the paper with some concluding remarks.

\section{Entanglement}
\label{ent}

Talking about entanglement is essential as it is a fundamental phenomenon that underpins much of QP non-intuitive and revolutionary aspects. Entanglement describes the deep and inherent correlations that may be established between quantum systems, challenging classical notions of locality and realism. It plays a crucial role in both theoretical foundations and practical applications of quantum physics, making it a topic of significant interest and importance.
Erwin Schroedinger, regarded entanglement as the {\em characteristic trait of quantum mechanics, the one that enforces its entire departure from classical lines of thought} \cite{schroedinger}. This statement underscores the unique nature of entanglement, where {\mga physical systems} become so deeply connected that the state of one cannot be described independently of the state of the other.

{\mga A bipartite entangled state cannot be created through local operations on the two parts 
alone, even if the two experimenters (say, Alice and Bob) are allowed to communicate. For 
entanglement to occur, the two systems must physically interact, resulting in a combined 
system that can only be described by a single wave function describing the properties of 
the entangled pair. This means that the concept of an independent 'state' for each subsystem 
loses significance, as there is no way to describe the properties of one subsystem without 
reference to the entire system. Consequently, observing the outcome of a measurement on 
one subsystem (Alice's) allows for a precise prediction of the corresponding measurement 
outcome on the other (Bob's)
\footnote{Notice that neither in the exhibition, nor in this paper, we discussed the difference between entanglement  (the property of quantum state of being not preparable by local operations and classical communication) and the stronger notion of Bell-nonlocality (the property of a state to lead to violation of BI with a suitable set of measurements performed at two distant sites).}.}
Entangled states are mathematically described by a non-factorizable superpositions of different classical options for the state of multipartite systems. For a bipartite system, the overall state  is entangled if it cannot be expressed as the tensor product of two states describing the two sub-systems independently. 

To highlight the structure of entangled states, let us consider one of the maximally-entangled Bell's states
\begin{equation}\label{bell_1}
|\Psi^-_{A,B}\rangle=\frac{1}{\sqrt{2}}\left(|0\rangle_A|1\rangle_B-|1\rangle_A|0\rangle_B\right) \ ,  
\end{equation}
which can be viewed as a superposition of the two classical configurations for the bipartite states $|0\rangle_A|0\rangle_B$ and $|1\rangle_A|1\rangle_B$. The state in Eq.~\ref{bell_1}) cannot be written as the product of the states describing Alice's and Bob's sub-systems separately, that is $|\Psi^-_{A,B}\rangle\neq \left(a|0\rangle_A + b|1\rangle_A\right) \left(c|0\rangle_B+d|1\rangle_B\right)$ for any choice of the coefficients $a,b,c$ and $d$.
We note that the structure of the entangled state in Eq.~(\ref{bell_1}) implies that the output values of a measurement on system A are perfectly anti-correlated with those of the same measurement on system B.
The state remains entangled also when the basis in both subsystems are rotated by an angle $\theta$. In fact, by defining: $|0_\theta\rangle=\cos\theta|0\rangle+\sin\theta|1\rangle$ and $|1_\theta\rangle=-\sin\theta|0\rangle+\cos\theta|1\rangle$, we get
\begin{equation}\label{bell_2}
|\Psi^-_{A,B}\rangle=\frac{1}{\sqrt{2}}\left(|0_\theta\rangle_A|1_\theta\rangle_B-|1_\theta\rangle_A|0_\theta\rangle_B\right) \ .
\end{equation}
This property is crucial for the discussion of the EPR paradox (see below).
{\mga An important point to emphasize here is that the local states (\MB measured by the two independent researchers} are mixed states in any basis: if Alice measures her state independently of Bob, the observable phenomenology indicates a maximally mixed state
\begin{equation}\label{mixed_1}
\rho_{A}=\frac{1}{2} \left(|0\rangle_{A A}\langle 0|+|1\rangle_{A A}\langle 1|\right) \ .
\end{equation}
In any rigorous dissemination program, it is essential to clarify the distinction between this mixed state and the “corresponding” superposition state,  $\psi_{A}=\left(|0\rangle_A-|1\rangle_A\right)/\sqrt{2}$. The mixed state expresses our knowledge that the state is in either 
$ | 0\rangle_{A}$  {\em or} $ | 1\rangle_{A}$ whereas the superposition conveys the more challenging notion of the system existing simultaneously in both  $ | 0\rangle_{A}$ {\em and}  $ | 1\rangle_{A}$.

In trying to communicate the concept of entanglement in an easily-understandable way, it is quite common to dwell on  the idea of perfect correlation (or, equivalently, anti-correlation), thus conveying the profoundly misleading message that  entangled states are nonclassical because they are perfectly correlated. Indeed, this is plain wrong, as also in classical physics we can produce mixed states that are perfectly correlated to each other in some variables: {\it e.g.} we could generate pairs of particles having the same color, the same size and the same weight, so that when Alice observes one property on her part of the system, she can perfectly forecast the output of Bob's measurement of the same observable.  There is no {\em quantumness} in this result and the emphasis on the perfect correlations within entangled states can be misleading. What is dramatically different for entangled state is that we can also change the measurement and observe a different property, which was not prepared before, and still observe perfect correlations. Moreover, those new observables could also be non-commuting with the others, as pointed out by EPR argument.
%%%
\section{
Correlation and Bell's Inequalities }
\label{clas_corr_Bell}

The standard way BI are usually  introduced {\VDR in textbooks}, which is the one probed in Nobel experiments, is the CHSH  formulation \cite{chsh}.
To understand how the CHSH inequality works, we can think of an experiment in which two people, Alice and Bob, who are in two laboratories far apart, each measure one of two parts of a system produced by some physical mechanism. 
%The two parts of the system can be related or unrelated. 
Each {\VDR  experimenter} can make measurements on their system: Alice can choose between two possible experiments (“$A_1$” and “$A_2$”), as can Bob (“$B_1$” and “$B_2$”). 
{\VDR The measured quantities are assumed to be dichotomic, so that } the result of each individual measurement can be $+1$ or $-1$. 
Alice and Bob randomly and independently choose which measurement to take and record the value obtained for each repetition of the experiment. There are four possibilities for the product of the measurement values: $A_1\cdot B_1$, $A_1\cdot B_2$, $A_2\cdot B_1$, $A_2\cdot B_2$, each of which has value $+1$ or $-1$.
Alice and Bob build correlation functions by calculating the average of the products of the 
results of all pairs of measurements taken on a large numbers of repetitions of the experiments, 
namely $\langle A_1\cdot B_1\rangle$, $\langle A_1\cdot B_2\rangle$, $\langle A_2\cdot B_1\rangle$, $\langle A_2\cdot B_2\rangle$, each having a value in the interval $[-1,1]$.
CHSH define the quantity 
\begin{equation}\label{eq:S}
    S=A_1\cdot  B_1+ A_1\cdot B_2 + A_2\cdot  B_1 -A_2\cdot B_2
\end{equation}
that can only be $S=\pm 2$ {for each pair}. By taking the average of Eq.~(\ref{eq:S}) one obtains
\begin{equation}\label{eq:chsh}
   \langle S\rangle=\langle A_1\cdot  B_1\rangle+ \langle A_1\cdot B_2\rangle +\langle A_2\cdot  B_1\rangle-\langle A_2\cdot B_2\rangle
\end{equation}
The CHSH inequality is a theorem that, under the assumptions of local realism and the possible existence of hidden variables, establishes the bound $|\langle S\rangle| \le 2$ (out of a possible range of $|\langle S\rangle| \le 4$ \cite{Popescu2014}). The two limiting cases, $\langle S\rangle 
= \pm 2$, correspond to systems that are perfectly correlated and anti-correlated, respectively. 
This result is general and rests on the assumptions of realism and locality. Specifically, these hypotheses are implicitly invoked when we assign to each measurement outcome $A_i$ and $B_i$ (for $i=1,2$) a well-defined, fixed value ($\pm 1$), independent of all other measurements. Bell’s inequalities (BI) therefore hold true for all classically correlated systems.

As mentioned above, this results holds true also if hidden variables are present \cite{Bell}, 
as long as realism and locality are considered. When moving from classical to quantum systems, 
things change drastically.  In fact, when  the maximally entangled Bell's states  described in 
Eq.\ref{bell_1} are considered, QP predicts the violation of CHSH inequality, with $|S|$ 
becoming as large as $2\sqrt{2}$ \cite{tsirelson}. This fundamental result is therefore 
in contradiction with the hypothesis of realism and locality and testing it experimentally 
is the route  Bell indicated to probe QP completeness.   
%%%
\section{Understanding the consequences of entanglement: the EPR paradox and Bell’s inequalities for all audiences}\label{story}
{\VDR In this section, we describe the educational path we have been using to introduce first the idea of entangled state and then the EPR argument, in a few informal and formal environments. 
To introduce the concept of entanglement, we discuss the case of the fundamental two-electron state of Helium (He) atom, which corresponds to two electrons with antiparallel 1/2-spins - by virtue of the Pauli exclusion principle. 

In the exhibition, this system can be introduced quite naturally, having previously  presented and discussed  (i) atomic orbitals as examples of quantum states and (ii)  spin  and its properties as probed by the Stern-Gerlach experiment. Moreover, the idea of two antiparallel electrons  occupying the He 1s orbital is somewhat familiar from standard high-school chemistry curriculum.}
Since the spins of the two electrons are perfectly anti-correlated in any direction, the correct mathematical form of the overall two-electron state is that of a singlet state 
\begin{equation}
\label{singlet}
|\Psi^{He}\rangle=\frac{1}{\sqrt{2}}\left(|\uparrow\rangle_1|\downarrow\rangle_2
-|\downarrow\rangle_1|\uparrow\rangle_2\right) \ ,  
\end{equation}
which is indeed one of the Bell's states.  This example, demonstrating that entanglement is a true property of physical systems, allows us to explain the meaning of Eq.~(\ref{singlet}) and to explicitly emphasize that, as there is no privileged orientation in the atom, 
it must be true for all directions, as in Eq.~(\ref{bell_2}).

{\VDR In the second step, we fully describe} the EPR argument, pointing out the role played by entanglement. 
The system considered by EPR (in Bohm's formulation of the paradox \cite{bohm}) is that of two electrons having total spin $\overline S=0$ in the state in Eq.~(\ref{singlet}). The two electrons are spatially separated and sent  to two experimenters, Alice and Bob, who can measure the spin of the electron along different axes. Looking at the form of the state, it is easy to be convinced that when Alice measures the spin along one axis, the value of spin measured by Bob along the same axis is also defined with certainty. The result is independent of the distance between Alice and Bob.
In their discussion, in addition to the validity of quantum theory, EPR make two assumptions:
\begin{itemize}
\item[(i)] Realism: ``If we can predict with certainty the result of a measurement on a system 
without interacting with it in any way, the measurement must correspond to a real property'' \cite{EPR}. A layperson’s interpretation of this idea is that the properties of physical systems 
({\it i.e.}, the observables) hold definite values regardless of whether we measure them or not. 
In other words, ``the moon remains in the sky even if we are not observing it'' \cite{mermin}.

\item[(ii)] Locality: the information obtained from a measurement on one of two isolated systems cannot produce a real change in the other, that is, the measurements made by Bob cannot depend on those made by Alice if their distance is sufficiently large.
The result of any measurement performed at Bob's site cannot depend on any actions taken 
by Alice—such as reading the result of her measurement—if they are outside each other's 
light-cone (or, to avoid the concept of a light-cone, if they are far enough apart that 
they cannot communicate, even by the fastest channel allowed by relativity).
\end{itemize}
In the EPR experiment, given the value of Alice's measurement of a spin component, the 
value of the same component by Bob can be predicted with certainty without the need of 
any action by Bob. Thus, for the realism hypothesis, the value of Bob's electron spin 
corresponding to the outcome of Alice's measurement must refer to a real property. If 
now Bob makes a measurement of another spin component at a different angle on the same 
electron, after the measurement he comes to know with certainty the value of that component. Consequently, Bob knows with certainty the values of two spin components that are mutually  incompatible according to QP. This contradiction is known as the EPR paradox.
We have three possible exits from the paradox: 
\begin{itemize}
\item[(i)] realism does not hold, {\it  i.e.}, it is not true that there are elements 
of reality pre-existing the measurements; 
\item[(ii)] locality does not hold, {\it i.e.}, the measurement on one part of the entangled state instantaneously determines the change in the state of the other part; 
\item[(iii)] QP is incomplete. It is well known that EPR's preference was for this third solution.
\end{itemize}
{\mga This discussion is perfectly understandable  by non experts, provided that 
the mathematical characterization of entangled states is introduced,  i.e., 
Eq.~(\ref{singlet}) is discussed. BI are then introduced as a quantitative translation 
of EPR argument in terms of correlations and are in fact based on the same assumptions 
of realism and locality, possibly complemented by the existence of hidden variables, 
whose role in their formulation must be carefully discussed. In this framework, in 
order to correctly communicate the meaning and  relevance of Eq.~(\ref{eq:chsh}), 
two key concepts need to be carefully addressed:
\begin{itemize}
\item[(i)] 
%The precise interpretation of the realism and locality assumptions, particularly their non-trivial nature. 
 The fact that realism and locality assumptions can not actually be taken for granted as `natural'.
While Einstein-Podolsky-Rosen were partially justified in arguing that any complete physical theory should satisfy both realist requirements (as exemplified by the `\textit{moon}' argument) and local causality constraints (rejecting ``spooky action at a distance''), their conclusions ultimately proved incompatible with experimental evidence.
\item[(ii)] 
The quantitative characterization of statistical correlation and its fundamental distinction from causation. Bell's crucial insight was recognizing that correlations provide a mathematical framework to test the EPR argument through experimental predictions. Moreover, since statistical descriptions inherently represent incomplete information, it should be clear that observed correlations between variables cannot be interpreted as causal relationships.  A canonical example illustrates this principle: while sunburn incidence and icecream consumption exhibit strong seasonal correlation, consuming icecream does not cause sunburns. Both phenomena instead share a common causal factor – increased solar radiation during hot months.
\end{itemize}
Within this framework, we examined the transition from classical correlation to Bell Inequality (BI) violation and its implications. We began by considering the differences between probabilistic predictions in classical and quantum physics. In classical physics, a probabilistic description is required only when information about the system is incomplete — namely, in the presence of unknown (hidden) variables. In contrast, quantum mechanics posits that the outcomes of an experiment are probabilistically distributed even when complete information about the system state is available. This fundamental discrepancy with classical predictions is essential for understanding the foundations of QP and can be clarified through the example of BI and their experimental violation. This, along with the supposed 'innocence' of the assumptions (realism and locality) and the simplicity of the algebra involved, makes BI particularly well-suited for introducing key aspects of QP.

It is important to note that while BI were originally introduced to address the EPR discussion, they are not a theorem exclusively about QP. Rather, they apply to any theory that seeks to describe nature, providing a means to determine whether such a theory can be made local-realist by incorporating hidden variables. In other words, BI allow us to pose a fundamental question to nature itself: `Is there an ultimate local-realist theory that describes you?'}
%%%
\section{Understanding CHSH inequality with cards}\label{cards}
{\mga Communicating BI to the general public poses a clear challenge due to their mathematical nature. In the probability-based formulation, a straightforward demonstration of BI validity using Venn diagrams has been proposed in several papers \cite{despagnat, chshmaccone}. On the other hand, the CHSH formulation of BI presented in Sect. 3 is  abstract, with the definition of $\langle S \rangle$ appearing somewhat ad hoc. Nevertheless, we believe this formulation is valuable in an educational context for two main reasons: (i) experimental evidence of BI violation ({\it e.g.}, Nobel Prize-winning experiments) directly refers to the CHSH formulation; and (ii) the CHSH approach explicitly introduces and calculates correlations, allowing us to clearly highlight the crucial differences between classical and quantum correlations in entangled states. To support this, we designed a simple `card game' to illustrate CHSH inequalities and help audiences understand their meaning and significance using only basic math.}
 
\begin{figure}
	\centering 
	\includegraphics[width=0.95\textwidth]{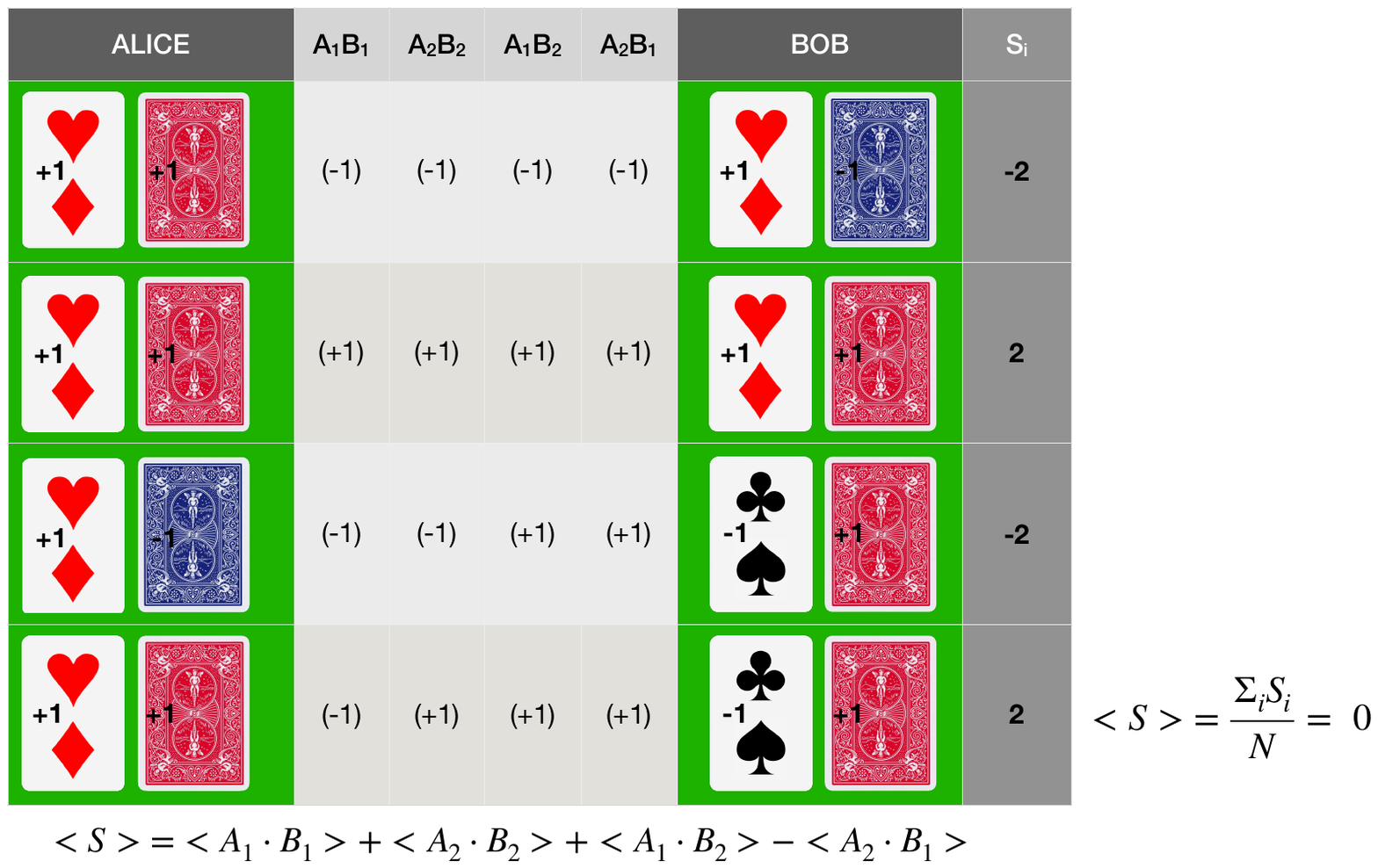}	
	\caption{ In the CHSH game two decks of cards are given to Alice and Bob, respectively. They measure both  properties of each card, {\it i.e.} the color of the back $A_1$,$B_1$, and the color of the front $A_2$,$B_2$.  A value of ($+1$) is associated to red back, ($-1$) to blue backs. Analogously,  (+1) corresponds to red suits (heart \& diamonds) and  ($-1$) to black ones (spades \& clubs). For each $i$-th pair of cards  they evaluate all terms $A_1\cdot A_2$, $A_1\cdot B_1$, etc) and calculate $S_i$. 
 As shown in the example, for every possible combination of cards $S_i$ is either $+2 $ or $-2$. The average $\langle S\rangle$ is therefore bound between $-2$ and $+2$, whatever the chosen ensemble. 
 The value $\langle S\rangle=-2$ corresponds to perfectly anti-correlated decks, while we obtain $\langle S\rangle=+2$ for perfectly correlated decks. Total randomness will result in $\langle S\rangle=0$, as a result of 
 equal number of $S_i=+2$ and $-2$.}
	\label{fig:chsh_cards}%
\end{figure}

As shown in Fig.~\ref{fig:chsh_cards}, the objects to be measured  are playing cards, characterized by two dichotomous properties: the color of their back (blue/red) and that of their front (black for spades and clubs, red for hearts and diamonds).  As the system is classic,  both properties can be measured for each card, the outcome of each measurement being $\pm 1$ (blu~$=-1$, red~$=1$; spades and clubs $=-1$, hearts and diamonds $=1$).
The dealer (Charlie) prepares two decks of $N$ cards  and gives the players  (Alice and Bob) one deck of cards each. In this way, the two decks can be seen as formed by $N$ pairs of cards, characterized by two properties whose correlation we are interested in  establishing and quantifying. Each player starts to {\it measure}  the cards, being careful to maintain the given original order, that is, to preserve the possible correlation within each pair, and compiles a table as shown in Fig.~\ref{fig:nobel}, where the outcomes of all measurements are registered.  
It is important here to notice that being the system classical, we are able to measure all the quantities at the same time and calculate the value of $S_i$ with $i=1,..N$,  for each row -- {\it i.e.} each pair of cards. Though this could be considered a gimmick of the game - as  in the real experiment only the average $\langle S\rangle$ can be evaluated and has a precise statistical meaning -- it allows us to easily convince everybody of the validity of  CHSH inequality.  
Indeed, it is easy to recognize that  for each row, whatever the colors of the card  pair (see  for instance the examples reported in Fig.~\ref{fig:chsh_cards}), $S_i$  can  assume only the values $\pm 2$, this meaning  that its mean value $\langle S \rangle={\sum_i S_i}/{N}$  must be bound between $\pm 2$ as well, {\it i.e.} $|\langle S\rangle|\leq 2$. 

This simple  card game provides  an elementary way to show CHSH inequality, accessible to anyone who knows how to calculate mean  values.  Moreover, it allows to naturally introduce the idea of perfect correlation between the properties of the two ensembles ($\langle S\rangle =2$, when the cards of each pair are identical), as well as perfect anticorrelation ($\langle S\rangle=-2$, when all the cards in each pair have opposite values) and all the cases in between (as for instance $\langle S\rangle=0$ when no correlation at all is present).  
In presenting this game, a few crucial aspects should be highlighted and discussed:
\begin{itemize}
\item[(i)] perfect correlation (anti-correlation) does exist in classical systems; it is not a prerogative of quantum states, nor of  entangled ones.

\item[(ii)]However we prepare the system, as long as the properties of the cards are ``real",  {\it i.e.} {\VDR we are dealing with the tangible, macroscopic cards, }
%the public hold in their hands,
there is no way BI can be violated.  

\item[(iii)] A crucial difference between the classical CHSH game  and the quantum one is that for the latter it is not possible to {\mga jointly} measure all properties, {\it i.e.} $S_i$ cannot be evaluated. This difference should be clearly stated, making it explicit  that  calculating  $S_i$ for each pair was a gimmick of the game to simplify the math.
In a true experiment (either classical or quantum), the value of $\langle S\rangle$ is  determined summing  all the correlation functions by calculating the average of the products of the results of all pairs of measurements, {\it i.e.}  by averaging each column of the table, rather than averaging over $S_i$. As long as the number of measurements is large enough ({\it i.e.} according to the Law of Large Numbers), the two procedures are equivalent and should lead to the same result.  This emphasizes the  statistical nature of the experimental result, clarifying the importance of collecting a large number of data, and of avoiding any possible bias in the choice of the quantities to be measured.

\item[(iv)] BI holds true also in the hypothesis of hidden variables, if realism and locality are assumed. Though the mathematical proof of  this  statement, which is  a major achievement of Bell's work \cite{Bell}, is more complex, it could and should be nevertheless clearly stated and explained.
\end{itemize}

The discussion of the results of the CHSH game paves the way to the presentation and discussion of the Nobel Prize-winning experiments. 
In order to highlight the key issues of these experiments, we involve the public in the staging --  a sort of `living crib' -- of the experiment, which is described in details in the next section.

% \section{CHSH violation and the Nobel experiments}
\section{Staging the Nobel Prize-winning experiments}\label{stage}

Conducting (or even simulating) the Nobel prize-winning experiments with an entangled source in a educational context -- though in principle possible \cite{Beck,Mitchell,Thorlabs}  --  requires sophisticated equipment and long acquisition times, making it unfeasible for our exhibition. 
We therefore  decided to stage the experiment, directly involving the public in  playing its fundamental phases. At the very beginning we make it clear  that - as  we are obviously not dealing with true entangled objects --  the staging is meant to describe and understand the key phases of the  experiment and not to reproduce its results.

\begin{figure}
	\centering 
	\includegraphics[width=0.95\textwidth]{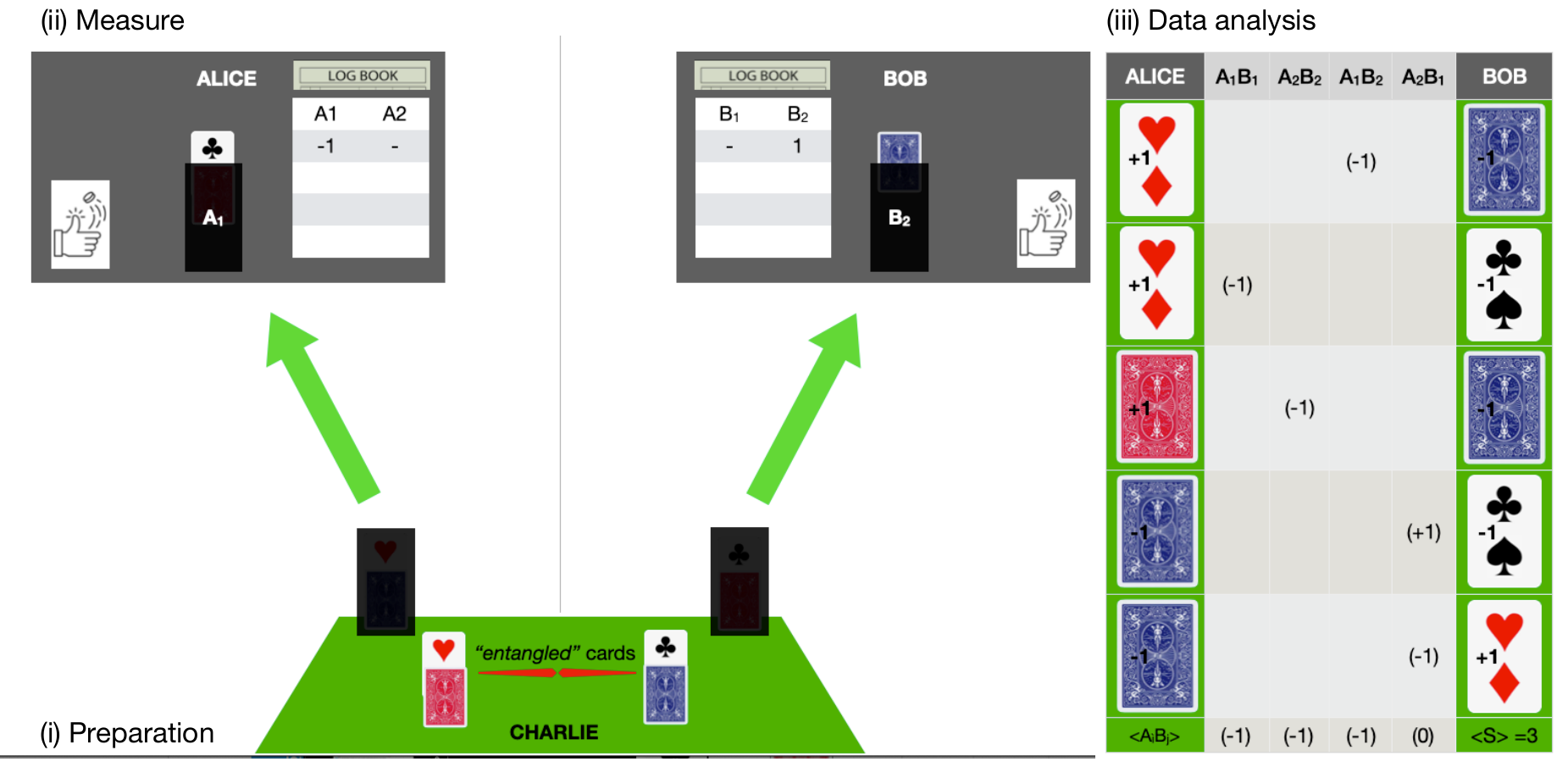}	
	\caption{Phases of the staged Nobel prize-winning experiment. (i) Charlie prepares pair of ``entangled cards'' and send them to Alice and Bob; (ii) Alice and Bob randomly chose and perform one measure; (iii) Charlie collect the measurement result and calculates $\langle S \rangle $. Three groups of  students play A, B and C, while the others bring the " entangled cards" from C to A and B.}
	\label{fig:nobel}%
\end{figure}

The entangled system is represented by  any  set of items characterized by two different dichotomous properties. In our case, we have used either cards as before, or small balls with different colors (green/yellow) and  solidity (soft/hard) (see Fig.~\ref{fig:chsh_balls}. The true experiment should be performed with two different pairs of non commuting observables, among which Alice and Bob should choose, {\it e.g.} photon polarizations at $0^o/45^o$ for Alice and $22.5^o/-22.5^o$ for Bob (see below the description of Aspect's experiment). Nevertheless, we decided to  keep it simple and used the same set of observables for both Alice and Bob.
The ``quantum'' nature of these object is staged by keeping them closed in small boxes,  so  that they cannot be seen until measured. Moreover, the boxes are designed so that only one property at a time can be sampled. In the case of the small balls for instance,  we can observe the color by peeking through a small hole, while we can feel their solidity with a finger, without looking into. In the case of cards, the box opening guarantees that only one side of the card (back or front) can be observed.  
As schematized in Fig.~\ref{fig:nobel}, panel a), the staged experiment proceed through the three  phases: 

\begin{figure}
	\centering 
	\includegraphics[width=0.95\textwidth]{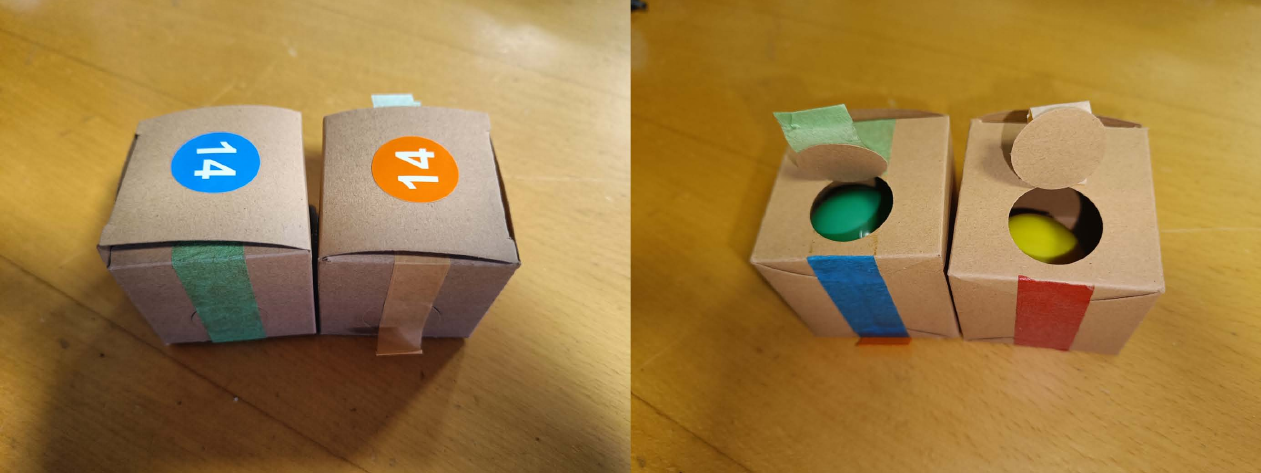}	
	\caption{In this version of the CHSH pairs of ping pong balls of different color and different solidity are prepared by Charlie and closed in two boxes with lateral apertures to perform the measurements. The boxes are then given to Alice and Bob who can just measure one property at a time: the color is observed by direct inspection while the solidity is tested with a finger. The measurement protocol follows that described in the caption of Fig.~\ref{fig:chsh_cards} .}
	\label{fig:chsh_balls}%
\end{figure}

\begin{itemize}
\item[(i)] Preparation of the entangled state: 
The person featuring Charlie  prepares  pairs of `entangled' items; as previously stated the items are not actually entangled, but are prepared so that each pair is anti-correlated, as shown in Fig.~\ref{fig:nobel} (i).  All possible combination of anti-correlated pairs are considered.
Once prepared, Charlie puts each item, separately, in a different closed box and sends them to Alice's and Bob's labs (which are located at two different tables). 

\item[(ii)]  Measure and data collection:
Alice and Bob, featured again by two people among the public, stay in their labs and receive their boxed item, which represent their part of the {\it entangled} system.
Independently and without communicating, they  randomly choose the type of measurement to be performed by throwing a coin. The importance of the true randomness and independence of their choices is highlighted, as it will  be recalled  afterwards, when the results of the actual  experiment are revised. 
As mentioned before, the boxes are designed so that only the chosen property can be measured, while the other remains unknown (undefined, in the quantum case). This procedure is repeated for a number of pairs, and each time Alice and Bob separately record the results of their measurements in a table in their logbook. At variance with the previous CHSH game, only half of the items in their tables will be filled, as shown in Fig.~\ref{fig:nobel} (ii). 

\item[(iii)] Data Analysis and evaluation of $\langle S\rangle$:
In the final step, Charlie collects both tables and calculates the value of $\langle S\rangle$. This is done by taking the average of each product column; on the average  only one fourth of the cells in each column are filled. At this stage, the  obtained value of $\langle S\rangle$ is not important (as the statistic is low, we could end up with whatever value!).  What is relevant is to show how the statistics is performed and  to  make it clear that by performing a large enough set of measurements, the  correlation values evaluated by averaging over a subset of the total possible pairs represents a statistically significant sample.  
\end{itemize}
 
 Having people acting the different phases of the experiment, even if only in a staged version of it, allows them to better understand the fundamental conceptual steps. 
 
Then, the real experiment can be discussed, both exploiting  computer simulations, such as {\it e.g.} the one available from the Flytrap project \cite{flytrapwebsite,flytrap}, and presenting one historical experiment, such as the one of the Nobel Prize Alain Aspect \cite{Aspect}.
Flytrap simulation of the experiment allows performing step-by step measurements and verifying the violation of BI (when statistics is high enough).
As to Aspect's experiment, performed in the early 1980s, the key points of the experimental setup are described. First of all, pairs of polarization-entangled photons are generated by a cascaded emission process in Calcium atoms, then the photons propagate to Alice's and Bob's laboratories where their polarization is measured. One of the key points of Aspect's setup is that Alice and Bob randomly choose the polarization measurement basis -- {\it i.e.} the orientation angle of the polarimeters -- during photon time of flight: this is to avoid any possible influence between the measurements in the two laboratories. Aspect's results for the values of $\langle S\rangle$ demonstrate a violation of the classical limit $\pm 2$ by 50 standard deviations. Overall, the experiments conducted by the Nobel Prize awardees have proven two key points:
\begin{itemize}
\item[(i)] BI are indeed violated, meaning that EPR’s assumption was incorrect and the `true theory of nature' cannot be  local-realistic, even with the inclusion of hidden variables.
\item[(ii)] The degree of violation aligns precisely with the predictions of quantum mechanics, indicating that, to the best of our current 
knowledge, QP is the most accurate theory describing nature.
\end{itemize}

\section{Path implementation and preliminary validation}\label{validation}
\mga{The formal methodological approach described above has been successfully implemented to teach entanglement, the EPR paradox, and Bell inequalities (BI) in diverse educational contexts, including:
\begin{itemize}
\item Quantum Technologies Summer Schools for 12th-grade students (Como, 2020--2024)
\item Online extracurricular courses for 13th-grade students \cite{MDPI}
\item Professional development programs for in-service high school teachers
\item Public outreach events through the Italian Quantum Weeks initiative. \cite{websiteIQWs}
\end{itemize} 
Notice that the educational framework outlined in Sect.~\ref{stage} formed the 
foundation of the 2023 public exhibition ``\textit{Dire l'indicibile}'' (``Speaking 
the Unspeakable''), presented in multiple Italian cities. Participants were guided 
through the exhibition by physics PhD students, with tours lasting approximately 90 minutes. While rigorous assessment of learning outcomes remains challenging in such an informal settings, we conducted participant surveys to gauge interest and satisfaction. Although not constituting formal validation, the results demonstrated strong public engagement and appreciation for the content. 
The successful maintenance of visitor attention throughout the abstract conceptual journey represents a significant achievement in science communication. Free-response comments revealed particular appreciation for:
\begin{itemize}
\item The guides' pedagogical skills in making complex concepts accessible
\item Interactive elements requiring active participant involvement.
\end{itemize}
These findings suggest that both facilitator expertise and hands-on engagement components are crucial success factors for science outreach initiatives.
A more structured implementation of our educational pathway was conducted through a three-day intensive program at the FIM Department in Modena, involving 26 12th-grade students (19 male, 7 female) from science-focused high schools (\textit{Liceo Scientifico} and \textit{Liceo delle Scienze Applicate}). Participant motivations included:
\begin{itemize}
\item Specific interest in physics (14 students)
\item General interest in science (10 students)
\item Non-academic motivations (3 students)
\end{itemize}
}

An introduction to QP is actually part of the fifth year (13th grade) of Italian  high school scientific curriculum and it usually follows an historical approach, introducing the old theory of quanta (Plank's equation, Bohr's atom, photoelectric effect). 
\begin{figure}[h!]
	\centering 
	\includegraphics[width=0.92\textwidth]{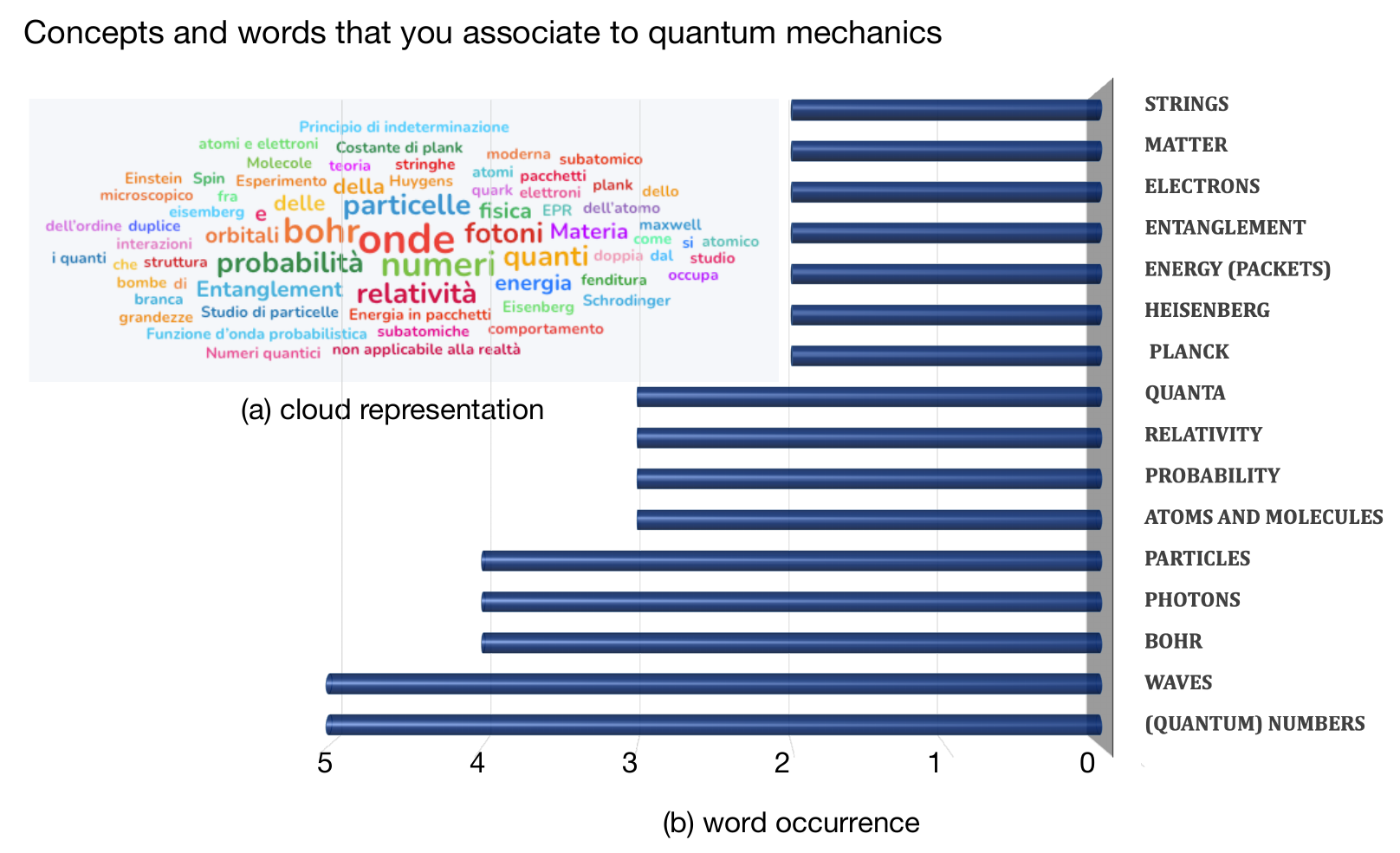}	
	\caption{Words and sentences that pupils associate to quantum physics, visualized as a cloud (in italian). The histogram shows the most used words (translated in english).  }
	\label{fig:nuvola}%
\end{figure}

As probed by informal inquiry, most of the pupils in 
their 12th year had  no previous knowledge on the subject. In Fig.~\ref{fig:nuvola} (a), we report, as a cloud of words, the answers to the preliminary question ``which concepts and words do you associate with quantum mechanics?''. It is interesting to briefly comment on the most cited words and concepts. Among 55 answers, only three were structured sentences (``Branch of modern physics that deals with the study of matter and waves at the size of the atomic and subatomic order'',  ``Double slit experiment, atomic structure, orbitals, interactions between atoms, string theory, subatomics, quanta, waves'' but also ``not applicable to reality.''), ten just provided scientist' names  (Bohr, Einstein, Heisenberg, Schroedinger and Planck, but also Maxwell and Huygens!). 
As shown in Fig.~\ref{fig:nuvola} (b),  the most cited words were `numbers' (most likely referring to quantum numbers), `waves' and `particles', followed by `quanta' and `probability'. It is interesting to note that `relativity' is also indicated several times. `Entanglement' is mentioned twice and EPR once.  When asked, only one pupil was able to tell what he knew about entanglement: ``two entangles particles - even when farthest apart – remain in relation to each other''.     
\begin{table}[h!]
    \centering
    \begin{tabular}{ll}
\hline\hline
& \\
 Q1  sentences  & \\
& \\
    a.  Bell inequalities are always verified in nature & $\times$\\
    b.  Bell inequalities are always verified under the assumption of realism and locality & $\checkmark$\\
    c.  Bell inequalities are always verified under the assumption of realism and non-locality &  $\times$\\
    d.  Bell inequalities violates the laws of relativity & $\times$\\
    e.   Bell inequalities  are always violates in quantum systems  & $\times$\\
    f.  Bell inequalities  are violated in some quantum systems & $\checkmark$\\
& \\
 \hline\hline
& \\
Q2   sentences  & \\
& \\
A.  Charlie produces a pair of entangled photons & $\checkmark$ \\
B. Charlie measures the photons &  $\times$\\
C.  Alice e Bob agree on which polarisation direction  should be measured &  $\times$\\
D.  Charlie sends  one photon to Alice and the other to Bob & $\checkmark$\\
E.  Charlie each time choose randomly weather to send  a photon to Alice or to Bob &  $\times$\\
F.  Alice receives a photon of the pair,  Bob the other & $\checkmark$\\
G.  Alice and Bob measure along the same polarisation directions  and write down the  results &  $\times$ \\
H.  Both Alice e Bob measure their photon,  only  along one polarisation direction &  \\ (chosen randomly) and write down the result & $\checkmark$\\
I.  Alice e Bob measure all polarisation direction of each photon &  $\times$\\
L.  Charlie produces a pair of \underline {quantum photons} & $\#$\\
M.  Alice receives both photons and than she sends them to Bob & $\times$\\
N.  Alice e Bob measure \underline {the polarisation} of their photon & $\#$\\
O.  Charlie produces one single entagled photon at a time &  $\times$\\
P.  Alice e Bob choose the polarisation direction to measure randomly and independently & $\checkmark$ \\
Q. Alice e Bob measurements are not commutative  & $\checkmark$\\
R. Alice e Bob measurements are commutative  &  $\times$   \\
& \\
\hline\hline
\end{tabular}
\caption{List of sentences of question (Q1) and (Q2). Correct (incorrect) answers are indicated with $\checkmark$ ($\times$) symbols.
    Two sentences (indicated by $\#$) in (Q2) present subtle (minor) flaws: in sentence N, the term \underline {``the polarization''} is used, without specifying its direction, in sentence O ``\underline {quantum photons}'' are mentioned, suggesting that classical photons may also exist.  A summary of pupils' answers are reported in Fig.~\ref{fig:results}}.
    \label{tab:q2}
\end{table}

To gather information on the path effectiveness, at the end of the stage, pupils' understanding was probed with a post-test, consisting both in open and multiple-choice items. {\VDR Open questions regarded (a) the definition of entanglement, (b) the meaning of the hidden variables proposed by Einstein, and (c) the difference between correlation and causation. 
We recognize that answers to (a) are quite difficult to evaluate: while  none of the pupils was able  to provide a correct  and complete definition of entanglement, most of them provided at least partially correct answers. As there is significant arbitrariness in how these answers can be evaluated,  we  decided not to further pursue this analysis (we can provide the original answers to the interested reader upon request). 
On the other hand, as far as (b) is concerned, most answers (17/26) capture the key point, that is the fact that hidden variables would provide a deterministic explanation of the probabilistic nature of QP, (7/26) answers were considered incorrect, as they mainly focus on the fact that these variables are unknown, while (2/26) pupils did not answer. 
It is interesting to note that question (c), was answered correctly only by (10/26) pupils. Among them, only one pupils provided a novel example, different from the one proposed during the class. Among wrong answers, most of them (8/26) confuse  causality {\it (causalit\`a)} with  randomness {\it (casualit\`a)} , while (4/26) confuse the idea of spurious correlation with that of correlation without causation.  This result shows that this key concept is not trivial, and that (at least in Italy) special care in the use of the similar terms ({\it causalit\`a/casualit\`a}) should be always taken.}    
Concerning the multiple-choice part of the test, pupils were asked to chose among several statements
the ones that correctly describe BI (Q1) and the statements which describe the fundamental phases of the CHSH experiments (Q2),
 as  enlisted in Table~\ref{tab:q2} (more then one choice was possible for each question).  

\begin{figure}[h!]
	\centering 
	\includegraphics[width=0.95\textwidth]{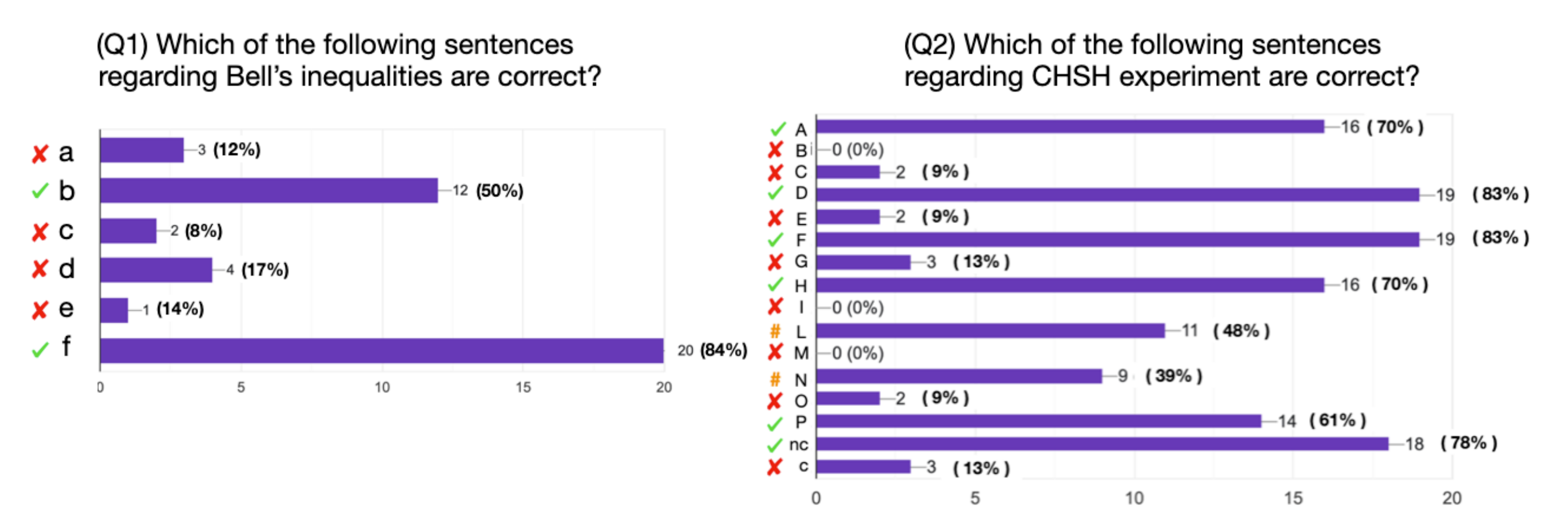}	
	\caption{(Q1) Among 26 pupils that answered to Q1 (see Table\ref{tab:q2} for the list of the sentences), 12 recognize as correct both sentence (b) and (f). Only four chose only wrong sentences, while two did not answer.  
 (Q2) Among 23 pupils that answered to Q2, 9 pupils selected all correct sentences, 7 made one mistake or fail to recognize one correct sentence, 2 pupils choose more than one wrong sentences, 5 fail to recognize more than one correct answer }
	\label{fig:results}%
\end{figure}

A summary of the  results of (Q1) and (Q2) answers is reported in Fig.~\ref{fig:results}. 
As far as the (Q1) question is concerned, the two correct sentences are chosen by the majority of pupils (10 (50\%) and 20 (83\%), respectively), while each of the four  wrong answers are chosen by less 4 students. It is interesting to note that the most chosen wrong answer (4 students) concerns an alleged ``violation of the laws of relativity''. While the sentence in itself is completely wrong and has also no logical meaning, this choice suggests -- as it can be expected -- that  QP non locality is the most difficult consequence of BI violation to be understood and accepted. 

Concerning  question (Q2), as detailed in the caption of Fig.~\ref{fig:results}, almost 80\% of the pupils  chose the correct sentences regarding Charlie, Alice and Bob roles (sentences A, D, F), the random choice of Alice and Bob measurements (H and P), and that the observables measured by A and B are not commutative.  Moreover,  more than 60\% of the pupils  selected all the correct answers or made only one mistake, while less than 20\% provided no answer or made several mistakes. 

\begin{figure}[h!]
	\centering 
	\includegraphics[width=0.97\textwidth]{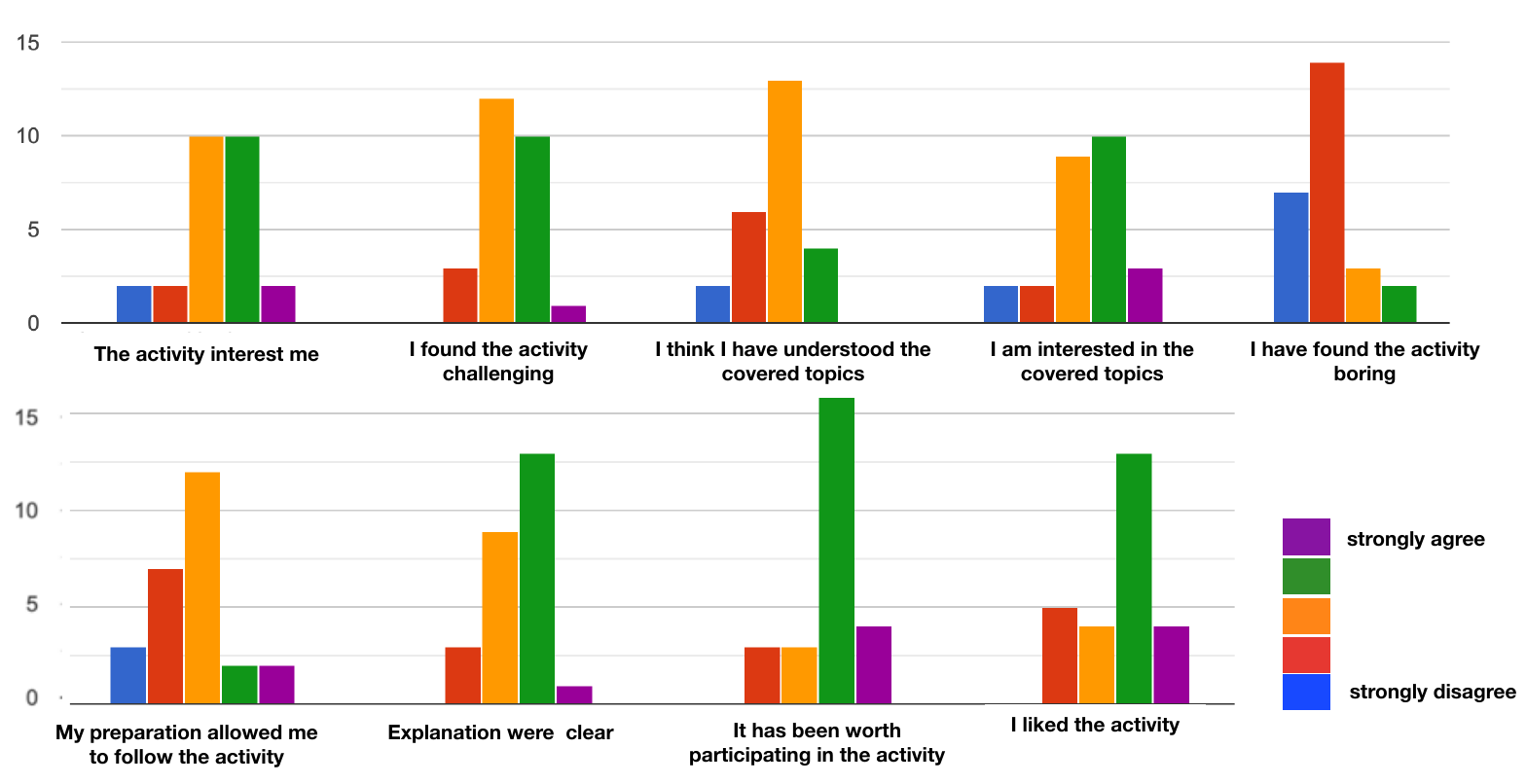}	
	\caption{ Results of the final feedback survey. Pupils were asked to (anonymously) state their agreement to the reported sentences} 
	\label{fig:survey}%
\end{figure}

Finally, participants were asked (Q3) to chronologically order the  phases they selected in (Q2). The results of this assessment component are summarized in Fig.~\ref{fig:results}. 
For Q3, we accepted two correct sequences:
\begin{itemize}
\item The basic sequence: A $\rightarrow$ D $\rightarrow$ F $\rightarrow$ Q $\rightarrow$ H
\item An extended sequence including additional phases L and O
\end{itemize}
Fifteen students correctly identified the temporal ordering, with phases H and Q 
considered chronologically interchangeable. While acknowledging limitations in sample 
size (26 participants) and assessment design (open-response questions with limited multiple-choice items), the outcomes appear positive and promising. These results 
suggest that:
\begin{itemize}
\item Motivated high school students can successfully engage with this complex subject matter
\item Meaningful learning outcomes can be achieved through targeted pedagogical approaches.
\end{itemize}
This interpretation is supported by the participant feedback results shown in Fig.~\ref{fig:survey}, which demonstrates strong overall satisfaction with the program. Particularly noteworthy  are the high scores in both assessment performance and positive survey responses, indicating both cognitive gains and affective engagement.
%%%
%%%
\section{Conclusion}\label{conc}
In recent years, with the emergence of quantum technologies, initiatives aimed at disseminating awareness about the foundations of Quantum Physics to the general public have become increasingly numerous. In this context, entanglement and its consequences are frequently used primarily to capture the public imagination, often waiving scientific 
rigor and sometimes leading to profound misconceptions. In fact, we believe that sharing the story of entanglement, the EPR paradox, and BI is essential, as these concepts are now woven into our cultural heritage and exemplify a landmark of scientific progress—one of the most remarkable achievements of twentieth-century physics. Furthermore, retracing the journey to experimentally verify BI violations over decades of precise, delicate work offers invaluable insights into the scientific process itself.

The real challenge lies in finding a way to explain the core character of this rather unique physical situation, so different from everyday experience, while avoiding misleading messages and misconceptions. Our experience has demonstrated that it is indeed possible to communicate these complex ideas to non-experts with only minimal use of mathematics, focusing on key elements like the structure of the entangled state and the derivation of BI. While a mathematical foundation is crucial, it must be paired with accessible, engaging communication that employs analogies and actively involves the audience.
Integrating ‘gaming’ activities proved to be a crucial factor in the success of our educational approach, fostering curiosity and a hands-on understanding that brought these abstract concepts to life. In particular, any educational path regarding entanglement and EPR should address the issue of effectively presenting the Nobel Prize-winning experiments that eventually provided evidence for BI violation. Reproducing the real experiments based on entangled photon sources, though in principle possible even with off-the-shelf apparatuses, is nevertheless beyond the possibilities of most educational initiatives. We showed that our staging activity is a valuable and viable solution to this problem. Besides the ease of implementation, this approach introduces personal experience and elements of embodied cognition,  facilitating grasping and retaining the key meaning of the otherwise quite cumbersome and complex experimental steps.

In conclusion, we believe that our strategy, which  uses of ad-hoc designed staging activities to convey reasoning, offers an effective and engaging way to 
communicate the profound implications of quantum entanglement and its experimental verification, making these concepts accessible to a broader audience.

\section*{Acknowledgements}
We thank Marco G. Genoni and Andrea Smirne for their contributions in the early stages of this work and for the insightful discussions. VDR and MGAP also thank Simone Cavazzoni, Gaia Forghieri, Giovanni Ragazzi, and Paolo Bordone for useful and entertaining discussions. This project has been supported by the University of Modena and Reggio Emilia, by the no-profit organization {\em Comitato Quantum}, and by the Department of Science and High Technology of Insubria University.

\section*{Conflicts of interest}
The authors declare no conflicts of interest.
\section*{Data availability}
Data are available from the authors upon reasonable request.
%%%
\bibliography{bibstaging}


\begin{thebibliography}{34}
\ifx \bisbn   \undefined \def \bisbn  #1{ISBN #1}\fi
\ifx \binits  \undefined \def \binits#1{#1}\fi
\ifx \bauthor  \undefined \def \bauthor#1{#1}\fi
\ifx \batitle  \undefined \def \batitle#1{#1}\fi
\ifx \bjtitle  \undefined \def \bjtitle#1{#1}\fi
\ifx \bvolume  \undefined \def \bvolume#1{\textbf{#1}}\fi
\ifx \byear  \undefined \def \byear#1{#1}\fi
\ifx \bissue  \undefined \def \bissue#1{#1}\fi
\ifx \bfpage  \undefined \def \bfpage#1{#1}\fi
\ifx \blpage  \undefined \def \blpage #1{#1}\fi
\ifx \burl  \undefined \def \burl#1{\textsf{#1}}\fi
\ifx \doiurl  \undefined \def \doiurl#1{\url{https://doi.org/#1}}\fi
\ifx \betal  \undefined \def \betal{\textit{et al.}}\fi
\ifx \binstitute  \undefined \def \binstitute#1{#1}\fi
\ifx \binstitutionaled  \undefined \def \binstitutionaled#1{#1}\fi
\ifx \bctitle  \undefined \def \bctitle#1{#1}\fi
\ifx \beditor  \undefined \def \beditor#1{#1}\fi
\ifx \bpublisher  \undefined \def \bpublisher#1{#1}\fi
\ifx \bbtitle  \undefined \def \bbtitle#1{#1}\fi
\ifx \bedition  \undefined \def \bedition#1{#1}\fi
\ifx \bseriesno  \undefined \def \bseriesno#1{#1}\fi
\ifx \blocation  \undefined \def \blocation#1{#1}\fi
\ifx \bsertitle  \undefined \def \bsertitle#1{#1}\fi
\ifx \bsnm \undefined \def \bsnm#1{#1}\fi
\ifx \bsuffix \undefined \def \bsuffix#1{#1}\fi
\ifx \bparticle \undefined \def \bparticle#1{#1}\fi
\ifx \barticle \undefined \def \barticle#1{#1}\fi
\bibcommenthead
\ifx \bconfdate \undefined \def \bconfdate #1{#1}\fi
\ifx \botherref \undefined \def \botherref #1{#1}\fi
\ifx \url \undefined \def \url#1{\textsf{#1}}\fi
\ifx \bchapter \undefined \def \bchapter#1{#1}\fi
\ifx \bbook \undefined \def \bbook#1{#1}\fi
\ifx \bcomment \undefined \def \bcomment#1{#1}\fi
\ifx \oauthor \undefined \def \oauthor#1{#1}\fi
\ifx \citeauthoryear \undefined \def \citeauthoryear#1{#1}\fi
\ifx \endbibitem  \undefined \def \endbibitem {}\fi
\ifx \bconflocation  \undefined \def \bconflocation#1{#1}\fi
\ifx \arxivurl  \undefined \def \arxivurl#1{\textsf{#1}}\fi
\csname PreBibitemsHook\endcsname

%%% 1
\bibitem[\protect\citeauthoryear{Einstein et~al.}{1935}]{EPR}
\begin{barticle}
\bauthor{\bsnm{Einstein}, \binits{A.}},
\bauthor{\bsnm{Podolsky}, \binits{B.}},
\bauthor{\bsnm{Rosen}, \binits{N.}}:
\batitle{Can quantum-mechanical description of physical reality be considered
  complete?}
\bjtitle{Phys. Rev.}
\bvolume{47},
\bfpage{777}--\blpage{780}
(\byear{1935})
\end{barticle}
\endbibitem

%%% 2
\bibitem[\protect\citeauthoryear{Bell}{1964}]{Bell}
\begin{barticle}
\bauthor{\bsnm{Bell}, \binits{J.S.}}:
\batitle{On the einstein podolsky rosen paradox}.
\bjtitle{Physics}
\bvolume{1}(\bissue{3}),
\bfpage{195}--\blpage{200}
(\byear{1964})
\end{barticle}
\endbibitem

%%% 3
\bibitem[\protect\citeauthoryear{Bell}{1987}]{speakable}
\begin{bbook}
\bauthor{\bsnm{Bell}, \binits{J.S.}}:
\bbtitle{Speakable and Unspeakable in Quantum Mechanics}.
\bpublisher{Cambridge University Press},
\blocation{Cambridge, UK}
(\byear{1987})
\end{bbook}
\endbibitem

%%% 4
\bibitem[\protect\citeauthoryear{Bohm}{1952a}]{hidden1}
\begin{barticle}
\bauthor{\bsnm{Bohm}, \binits{D.}}:
\batitle{A suggested interpretation of the quantum theory in terms of `hidden'
  variables. i}.
\bjtitle{Phys. Rev.}
\bvolume{85}(\bissue{2}),
\bfpage{166}--\blpage{179}
(\byear{1952})
\end{barticle}
\endbibitem

%%% 5
\bibitem[\protect\citeauthoryear{Bohm}{1952b}]{hidden2}
\begin{barticle}
\bauthor{\bsnm{Bohm}, \binits{D.}}:
\batitle{A suggested interpretation of the quantum theory in terms of `hidden'
  variables. ii}.
\bjtitle{Phys. Rev.}
\bvolume{85}(\bissue{2}),
\bfpage{180}--\blpage{193}
(\byear{1952})
\end{barticle}
\endbibitem

%%% 6
\bibitem[\protect\citeauthoryear{The Big Bell Test~Collaboration}{2018}]{BBT}
\begin{barticle}
\bauthor{\bsnm{The Big Bell Test~Collaboration}, \binits{.}}:
\batitle{Challenging local realism with human choices}.
\bjtitle{Nature}
\bvolume{557},
\bfpage{212}--\blpage{216}
(\byear{2018})
\end{barticle}
\endbibitem

%%% 7
\bibitem[\protect\citeauthoryear{}{}]{nobel}
\begin{botherref}
https://www.nobelprize.org/prizes/physics/2022/summary/
\end{botherref}
\endbibitem

%%% 8
\bibitem[\protect\citeauthoryear{}{}]{websiteIQWs}
\begin{botherref}
https://quantumweeks.it/
\end{botherref}
\endbibitem

%%% 9
\bibitem[\protect\citeauthoryear{De~Renzi et~al.}{}]{manu}
\begin{botherref}
\oauthor{\bsnm{De~Renzi}, \binits{V.}},
\oauthor{\bsnm{Paris}, \binits{M.G.A.}},
\oauthor{\bsnm{Bondani}, \binits{M.}}:
Manuscript in preparation.
\end{botherref}
\endbibitem

%%% 10
\bibitem[\protect\citeauthoryear{}{}]{QTEdu}
\begin{botherref}
https://qtedu.eu/
\end{botherref}
\endbibitem

%%% 11
\bibitem[\protect\citeauthoryear{Migda{\l}
  et~al.}{2022}]{migdal2022visualizing}
\begin{barticle}
\bauthor{\bsnm{Migda{\l}}, \binits{P.}},
\bauthor{\bsnm{Jankiewicz}, \binits{K.}},
\bauthor{\bsnm{Grabarz}, \binits{P.}},
\bauthor{\bsnm{Decaroli}, \binits{C.}},
\bauthor{\bsnm{Cochin}, \binits{P.}}:
\batitle{Visualizing quantum mechanics in an interactive simulation--virtual
  lab by quantum flytrap}.
\bjtitle{Optical Engineering}
\bvolume{61}(\bissue{8}),
\bfpage{081808}--\blpage{081808}
(\byear{2022})
\end{barticle}
\endbibitem

%%% 12
\bibitem[\protect\citeauthoryear{Bondani et~al.}{2022}]{MDPI}
\begin{barticle}
\bauthor{\bsnm{Bondani}, \binits{M.}},
\bauthor{\bsnm{Chiofalo}, \binits{M.L.}},
\bauthor{\bsnm{Ercolessi}, \binits{E.}},
\bauthor{\bsnm{Macchiavello}, \binits{C.}},
\bauthor{\bsnm{Malgieri}, \binits{M.}},
\bauthor{\bsnm{Michelini}, \binits{M.}},
\bauthor{\bsnm{Mishina}, \binits{O.}},
\bauthor{\bsnm{Onorato}, \binits{P.}},
\bauthor{\bsnm{Pallotta}, \binits{F.}},
\bauthor{\bsnm{Satanassi}, \binits{S.}},
\bauthor{\bsnm{Stefanel}, \binits{A.}},
\bauthor{\bsnm{Sutrini}, \binits{C.}},
\bauthor{\bsnm{Testa}, \binits{I.}},
\bauthor{\bsnm{Zuccarini}, \binits{G.}}:
\batitle{Introducing quantum technologies at secondary school level: Challenges
  and potential impact of an online extracurricular course}.
\bjtitle{Physics}
\bvolume{4}(\bissue{4}),
\bfpage{1150}--\blpage{1167}
(\byear{2022})
\end{barticle}
\endbibitem

%%% 13
\bibitem[\protect\citeauthoryear{Schneble et~al.}{2025}]{10.1119/5.0211535}
\begin{barticle}
\bauthor{\bsnm{Schneble}, \binits{D.}},
\bauthor{\bsnm{Wei}, \binits{T.-C.}},
\bauthor{\bsnm{Kelly}, \binits{A.M.}}:
\batitle{Quantum information science and technology high school outreach:
  Conceptual progression for introducing principles and programming skills}.
\bjtitle{American Journal of Physics}
\bvolume{93}(\bissue{1}),
\bfpage{88}--\blpage{97}
(\byear{2025})
\end{barticle}
\endbibitem

%%% 14
\bibitem[\protect\citeauthoryear{Brang et~al.}{2024}]{bitz}
\begin{barticle}
\bauthor{\bsnm{Brang}, \binits{M.}},
\bauthor{\bsnm{Franke}, \binits{H.}},
\bauthor{\bsnm{Greinert}, \binits{F.}},
\bauthor{\bsnm{Ubben}, \binits{M.S.}},
\bauthor{\bsnm{Hennig}, \binits{F.}},
\bauthor{\bsnm{Bitzenbauer}, \binits{P.}}:
\batitle{Spooky action at a distance? a two-phase study into learners’ views
  of quantum entanglement}.
\bjtitle{Proc. SPIE}
\bvolume{11},
\bfpage{33}
(\byear{2024})
\end{barticle}
\endbibitem

%%% 15
\bibitem[\protect\citeauthoryear{Giliberti et~al.}{2023}]{gili23}
\begin{barticle}
\bauthor{\bsnm{Giliberti}, \binits{M.A.L.}},
\bauthor{\bsnm{Lovisetti}, \binits{L.}},
\bauthor{\bsnm{Olivares}, \binits{S.}},
\bauthor{\bsnm{Paris}, \binits{M.G.A.}}:
\batitle{Meccanica quantistica, entanglement e nonlocalit{\`a}}.
\bjtitle{Giornale di Fisica}
\bvolume{64}(\bissue{2}),
\bfpage{161}--\blpage{183}
(\byear{2023})
\end{barticle}
\endbibitem

%%% 16
\bibitem[\protect\citeauthoryear{López-Incera and
  Dür}{2019}]{10.1119/1.5086275}
\begin{barticle}
\bauthor{\bsnm{López-Incera}, \binits{A.}},
\bauthor{\bsnm{Dür}, \binits{W.}}:
\batitle{Entangle me! a game to demonstrate the principles of quantum
  mechanics}.
\bjtitle{American Journal of Physics}
\bvolume{87}(\bissue{2}),
\bfpage{95}--\blpage{101}
(\byear{2019})
\end{barticle}
\endbibitem

%%% 17
\bibitem[\protect\citeauthoryear{López-Incera
  et~al.}{2020}]{Lopez-Incera_2020}
\begin{barticle}
\bauthor{\bsnm{López-Incera}, \binits{A.}},
\bauthor{\bsnm{Hartmann}, \binits{A.}},
\bauthor{\bsnm{Dür}, \binits{W.}}:
\batitle{Encrypt me! a game-based approach to bell inequalities and quantum
  cryptography}.
\bjtitle{European Journal of Physics}
\bvolume{41}(\bissue{6}),
\bfpage{065702}
(\byear{2020})
\doiurl{10.1088/1361-6404/ab9a67}
\end{barticle}
\endbibitem

%%% 18
\bibitem[\protect\citeauthoryear{Foti et~al.}{2021}]{Foti}
\begin{barticle}
\bauthor{\bsnm{Foti}, \binits{C.}},
\bauthor{\bsnm{Anttila}, \binits{D.}},
\bauthor{\bsnm{Maniscalco}, \binits{S.}},
\bauthor{\bsnm{Chiofalo}, \binits{M.L.}}:
\batitle{Quantum physics literacy aimed at k12 and the general public.}
\bjtitle{Universe}
\bvolume{7},
\bfpage{86}
(\byear{2021})
\end{barticle}
\endbibitem

%%% 19
\bibitem[\protect\citeauthoryear{}{}]{qplaylearn}
\begin{botherref}
https://qplaylearn.com/
\end{botherref}
\endbibitem

%%% 20
\bibitem[\protect\citeauthoryear{Bondani et~al.}{2024}]{QTris}
\begin{barticle}
\bauthor{\bsnm{Bondani}, \binits{M.}},
\bauthor{\bsnm{Caprara}, \binits{S.}},
\bauthor{\bsnm{Chiarello}, \binits{F.}},
\bauthor{\bsnm{Dabbicco}, \binits{M.}},
\bauthor{\bsnm{Hamma}, \binits{A.}},
\bauthor{\bsnm{Malgieri}, \binits{I.}},
\bauthor{\bsnm{Marzoli}, \binits{I.}},
\bauthor{\bsnm{Nazzaro}, \binits{M.}},
\bauthor{\bsnm{Paladino}, \binits{E.}}:
\batitle{Qtris: a quantum game}.
\bjtitle{Proc. SPIE}
\bvolume{12993}(\bissue{Quantum Technologies 2024}),
\bfpage{129930}
(\byear{2024})
\end{barticle}
\endbibitem

%%% 21
\bibitem[\protect\citeauthoryear{Schroedinger}{1935}]{schroedinger}
\begin{barticle}
\bauthor{\bsnm{Schroedinger}, \binits{E.}}:
\batitle{Discussion of probability relations between separated systems}.
\bjtitle{Mathematical Proceedings of the Cambridge Philosophical Society}
\bvolume{31},
\bfpage{555}--\blpage{563}
(\byear{1935})
\end{barticle}
\endbibitem

%%% 22
\bibitem[\protect\citeauthoryear{Clauser et~al.}{1969}]{chsh}
\begin{barticle}
\bauthor{\bsnm{Clauser}, \binits{J.F.}},
\bauthor{\bsnm{Horne}, \binits{M.A.}},
\bauthor{\bsnm{Shimony}, \binits{A.}},
\bauthor{\bsnm{Holt}, \binits{R.A.}}:
\batitle{Proposed experiment to test local hidden-variable theories}.
\bjtitle{Phys. Rev. Lett.}
\bvolume{23}(\bissue{15}),
\bfpage{880}--\blpage{884}
(\byear{1969})
\end{barticle}
\endbibitem

%%% 23
\bibitem[\protect\citeauthoryear{Popescu}{2014}]{Popescu2014}
\begin{barticle}
\bauthor{\bsnm{Popescu}, \binits{S.}}:
\batitle{Nonlocality beyond quantum mechanics}.
\bjtitle{Nature Physics}
\bvolume{10}(\bissue{4}),
\bfpage{264}--\blpage{270}
(\byear{2014})
\doiurl{10.1038/nphys2916}
\end{barticle}
\endbibitem

%%% 24
\bibitem[\protect\citeauthoryear{Cirel'son}{1980}]{tsirelson}
\begin{barticle}
\bauthor{\bsnm{Cirel'son}, \binits{B.S.}}:
\batitle{Quantum generalizations of bell's inequality}.
\bjtitle{Letters in Mathematical Physics}
\bvolume{4}(\bissue{2}),
\bfpage{93}--\blpage{100}
(\byear{1980})
\doiurl{10.1007/BF00417500}
\end{barticle}
\endbibitem

%%% 25
\bibitem[\protect\citeauthoryear{Bohm and Aharonov}{1957}]{bohm}
\begin{barticle}
\bauthor{\bsnm{Bohm}, \binits{D.}},
\bauthor{\bsnm{Aharonov}, \binits{Y.}}:
\batitle{Discussion of experimental proof for the paradox of einstein, rosen,
  and podolsky}.
\bjtitle{Phys. Rev.}
\bvolume{108}(\bissue{4}),
\bfpage{1070}--\blpage{1076}
(\byear{1957})
\end{barticle}
\endbibitem

%%% 26
\bibitem[\protect\citeauthoryear{Mermin}{1985}]{mermin}
\begin{barticle}
\bauthor{\bsnm{Mermin}, \binits{N.D.}}:
\batitle{Is the moon there when nobody looks? reality and the quantum theory}.
\bjtitle{Physis Today}
\bvolume{April},
\bfpage{38}--\blpage{47}
(\byear{1985})
\end{barticle}
\endbibitem

%%% 27
\bibitem[\protect\citeauthoryear{d’Espagnat}{1979}]{despagnat}
\begin{barticle}
\bauthor{\bsnm{d’Espagnat}, \binits{B.}}:
\batitle{The quantum theory and reality}.
\bjtitle{Scientific American}
\bvolume{241},
\bfpage{158}--\blpage{181}
(\byear{1979})
\end{barticle}
\endbibitem

%%% 28
\bibitem[\protect\citeauthoryear{Maccone}{2013}]{chshmaccone}
\begin{barticle}
\bauthor{\bsnm{Maccone}, \binits{L.}}:
\batitle{A simple proof of {B}ell's inequality}.
\bjtitle{Am. J. Phys.}
\bvolume{81},
\bfpage{854}--\blpage{859}
(\byear{2013})
\end{barticle}
\endbibitem

%%% 29
\bibitem[\protect\citeauthoryear{Beck and Dederick}{2014}]{Beck}
\begin{barticle}
\bauthor{\bsnm{Beck}, \binits{M.}},
\bauthor{\bsnm{Dederick}, \binits{E.}}:
\batitle{Quantum optics laboratories for undergraduates}.
\bjtitle{Proc. SPIE}
\bvolume{9289},
\bfpage{92891}
(\byear{2014})
\end{barticle}
\endbibitem

%%% 30
\bibitem[\protect\citeauthoryear{Dehlinger and Mitchell}{2002}]{Mitchell}
\begin{barticle}
\bauthor{\bsnm{Dehlinger}, \binits{D.}},
\bauthor{\bsnm{Mitchell}, \binits{M.W.}}:
\batitle{Entangled photons, nonlocality, and bell inequalities in the
  undergraduate laboratory}.
\bjtitle{Am. J. Phys.}
\bvolume{70},
\bfpage{903}--\blpage{910}
(\byear{2002})
\end{barticle}
\endbibitem

%%% 31
\bibitem[\protect\citeauthoryear{}{}]{Thorlabs}
\begin{botherref}
https://www.thorlabs.com/newgrouppage9.cfm?objectgroup$\_$id=15827
\end{botherref}
\endbibitem

%%% 32
\bibitem[\protect\citeauthoryear{}{}]{flytrapwebsite}
\begin{botherref}
https://lab.quantumflytrap.com/lab?mode=waves
\end{botherref}
\endbibitem

%%% 33
\bibitem[\protect\citeauthoryear{Migdal et~al.}{2022}]{flytrap}
\begin{barticle}
\bauthor{\bsnm{Migdal}, \binits{P.}},
\bauthor{\bsnm{Jankiewicz}, \binits{K.}},
\bauthor{\bsnm{Grabarz}, \binits{P.}},
\bauthor{\bsnm{Decaroli}, \binits{C.}},
\bauthor{\bsnm{Cochin}, \binits{P.}}:
\batitle{Visualizing quantum mechanics in an interactive simulation – virtual
  lab by quantum flytrap}.
\bjtitle{Optical Engineering}
\bvolume{61}(\bissue{8}),
\bfpage{081808}
(\byear{2022})
\end{barticle}
\endbibitem

%%% 34
\bibitem[\protect\citeauthoryear{Aspect et~al.}{1982}]{Aspect}
\begin{barticle}
\bauthor{\bsnm{Aspect}, \binits{A.}},
\bauthor{\bsnm{Grangier}, \binits{P.}},
\bauthor{\bsnm{Roger}, \binits{G.}}:
\batitle{Experimental realization of einstein-podolsky-rosen-bohm
  gedankenexperiment: A new violation of bell's inequalities}.
\bjtitle{Phys. Rev. Letters}
\bvolume{49},
\bfpage{91}--\blpage{94}
(\byear{1982})
\end{barticle}
\endbibitem

\end{thebibliography}
\end{document}